\documentclass{appolb}
\usepackage{graphicx}
\usepackage{color}
\usepackage[colorlinks]{hyperref}
\usepackage{amssymb}
\usepackage{amsmath}
\newcommand{\ar}{\arrowvert}
\newcommand{\ra}{\rangle}
\newcommand{\la}{\langle}

\newcommand{\cd}{\! \cdot \!}
\newcommand{\be}{\begin{equation}}
\newcommand{\ee}{\end{equation}}
\newcommand{\ba}{\begin{eqnarray}}
\newcommand{\ea}{\end{eqnarray}}

\begin{document}
\title{Triply heavy baryon mass estimated within pNRQCD
\thanks{FJLE thanks the organizers of Excited QCD 2013, Sarajevo, for the invitation to present this work, supported by grants FIS2008-01323, FPA2011-27853-01 and by the Austrian Science Fund (FWF) under project no. M1333-N16 (R.W).}
}
\author{Felipe J. Llanes-Estrada$^1$, Olga I. Pavlova$^1$,\\
 and Richard Williams$^2$
\address{$^1$ Dept. F\'{\i}sica Te\'orica I, Univ. Complutense de Madrid, 28040, Madrid, Spain\\
$^2$ Institut f\"ur Physik, K.-F.-Univ. Graz, Universit\"atsplatz 5, A-8010 Graz, Austria}
}
\maketitle
\begin{abstract}
Potential-NRQCD offers an effective-theory based approach to heavy-quark physics. While meson $Q\bar{Q}$ computations are tractable in pure $\alpha_s$-perturbation theory, more complex
many-body quark systems transcend it. A possibility inherited from nuclear physics is to employ the perturbative static potentials in a numerical diagonalization, eventually obtaining the exact lowest eigenvalue in each channel for a given-order perturbative potential. The power counting is manifest in the potential instead of the spectrum.
The NNLO-potential for the 3-body problem is already available, so we have addressed triply-heavy baryons in this initial work with a computer-aided 2-parameter variational treatment.
\end{abstract}

\section{pNRQCD 3-quark potentials, including intrinsic 3-body force}
\begin{figure}[htb]
\centerline{\includegraphics*[width=5cm]{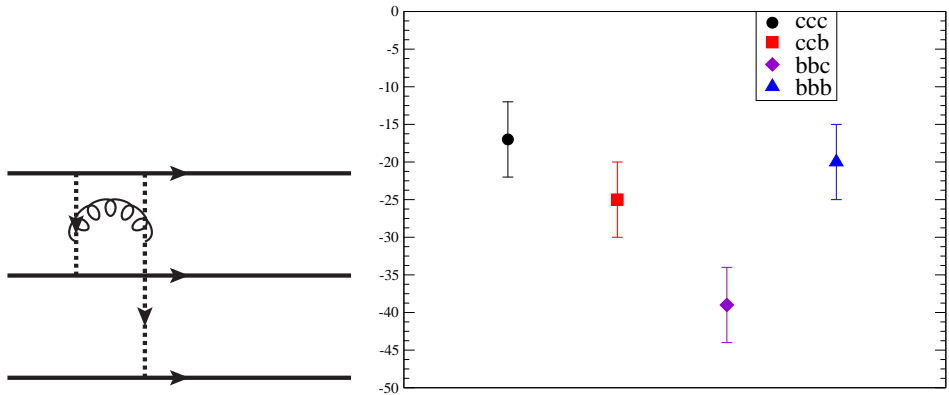}
\includegraphics*[width=7.5cm]{FIGS.DIR/Threebodymineonly.eps}}
\caption{Left: Feynman diagram yielding an intrinsic 3-quark force in a non-Abelian gauge theory. Right: computed mass difference due to the addition of such force. This is good news for Faddeev baryon calculations: at least in the heavy-quark limit we can state that 3-body forces are small~\cite{SanchisAlepuz:2011jn}.}
\label{Fig:3body}
\end{figure}

The LO potential of three heavy quarks in pNRQCD~\cite{Brambilla:2004jw} is a
($\Delta$-shaped) sum of 2-body Coulomb interactions dependent on pairwise coordinates,
\be \label{LOpot}
V^{(0)}_{LO} = \frac{-2\alpha_s}{3}\left(
\frac{1}{\ar {\bf r}_1-{\bf r}_2\ar}+
\frac{1}{\ar {\bf r}_2-{\bf r}_3\ar}+
\frac{1}{\ar {\bf r}_3-{\bf r}_1\ar} \right) \ .
\ee
The Hamiltonian resulting from adding the heavy-quark kinetic energy to this potential has already been variationally treated in the past~\cite{Jia}.
This potential is simple enough that analytical approximations to the ground-state binding energy are possible, but the situation changes drastically when adding NLO and NNLO potentials.
The full NLO and part of the NNLO ones are still $\Delta$-shaped and mimic the two-body potential in a meson
\ba
V^{(0)}_{LO}+ V^{(0)}_{NLO} &=&  
\frac{-2}{3} \sum_i \alpha_s(\ar {\bf r}_i \ar^{-2}) \frac{1}{\ar {\bf  r}_i\ar } \times 
 \left[ 1 + \frac{\alpha_s(\ar {\bf r}_i \ar^{-2})}{4\pi}\left(2\beta_0\gamma_E + a_1\right)\right] \\
V_{NNLO-2}^{(0)}  &=&  
\frac{-2}{3} \sum_i \frac{\alpha_s({\bf r}_i^{-2})}{\ar {\bf r}_i\ar} \frac{\alpha_s({\bf r}_i^{-2})^2}{(4\pi)^2}  \\  \nonumber  &\times& 
\left( a_2 -36\pi^2+3\pi^4 +\left( \frac{\pi^2}{3} + 4\gamma_E^2\right)
\beta_0^2 + \gamma_E(4a_1\beta_0+2\beta_1)
\right),
\ea
but at NNLO an intrinsic 3-body part is also present. It is a consequence of Yang--Mills theory being non-Abelian, thus featuring a three-gluon vertex: the Feynman diagram is depicted in the left-plot of figure~\ref{Fig:3body}. This intrinsic 3-body force is simplest in momentum space (a fast Fourier algorithm easily transforms potentials between momentum and coordinate representations),
\be \label{NNLO3}
\hat{V}_{\rm NNLO-3}^{(0)}= \frac{(-i/2)(4\pi)^3 \alpha_s^3}
{8\ar {\bf q}_2\ar^2 \ar{\bf q}_3\ar^2} 
\left[
\frac{\ar {\bf q}_2+{\bf q}_3\ar}{\ar {\bf q}_2 \ar\ar {\bf q}_3 \ar}
+\frac{{\bf q}_2\cd {\bf q}_3 + \ar {\bf q}_2 \ar\ar {\bf q}_3 \ar}
{\ar {\bf q}_2 \ar\ar {\bf q}_3 \ar \ar {\bf q}_2+{\bf q}_3\ar}
-\frac{1}{\ar {\bf q}_2 \ar}  -\frac{1}{\ar {\bf q}_3 \ar}  
\right]
\ee

We also advance in the figure (right plot) our numerical estimate of this  3-body force on the $\Omega_{QQQ}$ $J^P=3/2^+$ baryons for $QQQ=ccc,ccb,bbc,bbb$. We found it to be of order 30-40 MeV by computing the spectrum with and without $\hat{V}_{\rm NNLO-3}^{(0)}$ from Eq.~(\ref{NNLO3}). We appraise the effect as too small for unambiguous theory to extract it from the ground-state spectrum alone. 

\begin{figure}[htb]
\centerline{\includegraphics*[width=8cm]{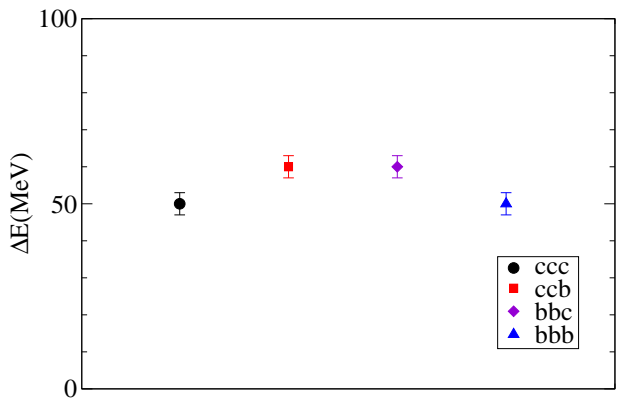}}
\caption{Triply heavy-baryon mass shifts due to the subleading $1/m_Q$ potential. }
\label{Fig:1overm}
\end{figure}

In addition to the static potential, we have also considered the first $1/m$ correction, that to this order is also the sum of 2-body potentials~\cite{Kniehl:2001ju}
\begin{align}
V_{m^{-1}} &= -\frac{7}{9}\frac{\alpha_s^2(\mu)}{m_r r^2} 
\label{1overmpot} 
&-\frac{\alpha_s^3(\mu)}{3\pi m_r r^2}
\left\{
-b_2 +  \log(e^{2\gamma_E}\mu^2 r^2)
\left( \frac{7\beta_0}{6}+\frac{68}{3} \right) \right\}\ .
\end{align}
Its effect on the triply heavy baryons, shown in figure~\ref{Fig:1overm} is moderate.

\section{Variational, numerical treatment of the 3-body problem}
The three-body (and more so the many-body) problem is not analytically tractable with  non-trivial potentials. We therefore resort to a numerical, variational treatment. This is in the spirit of modern nuclear physics~\cite{Weinberg:1991um} where scattering observables or spectra require solving a many-body equation, and the perturbative counting is manifest in the underlying potential.
Before embarking into baryons, we worked out several meson observables within the same framework, but we~\cite{LlanesEstrada:2011kc} and others~\cite{Laschka:2011zr} already documented these methods. Here we just comment briefly on the 3-body problem.

In the spirit of~\cite{LlanesEstrada:2011jd,Bicudo:2009cr}, we proceed variationally  with a simple wave-function ansatz.
This is used with the pNRQCD Hamiltonian through the Rayleigh-Ritz variational principle, an upper bound to the binding energy,
\begin{align}
\frac{\la \psi_{\alpha_\rho  \alpha_\lambda} \ar H_{pNRQCD}\ar \psi_{\alpha_\rho  \alpha_\lambda} \ra}{\la \psi_{\alpha_\rho  \alpha_\lambda} \ar \psi_{\alpha_\rho  \alpha_\lambda} \ra} \ge E_0 
\end{align}

We vary the two parameters  $\alpha_\rho$, $\alpha_\lambda$ to
find the optimum energy upper-bound for the  ansatz. These
two parameters control the momentum-space spread in the 3-body Jacobi-coordinates, that for a hadron at rest read
\begin{align}
k_\rho \equiv \frac{k_1-k_2}{\sqrt{2}} \  \ \  \
k_{\lambda} \equiv \sqrt{\frac{3}{2}}(k_1+k_2) \ \ \  \
k_3 \equiv -k_1-k_2 \ .
\end{align}
We choose as ansatz $\psi(k_\rho,k_\lambda)_{\alpha_\rho \alpha_\lambda} = Y_{00}(k_\rho) Y_{00}(k_{\lambda})
e^{-k_{\rho}/\alpha_\rho-k_\lambda/\alpha_{\lambda}}$,
although we have also checked other forms such as a rational function, with consistent results. The error incurred in this variational approximation
is estimated by employing the same technique for three atomic-physics systems (orthohelium, parahelium, and the dihydrogen-cation).

\begin{figure}[htb]
\centerline{
\includegraphics*[width=8.5cm]{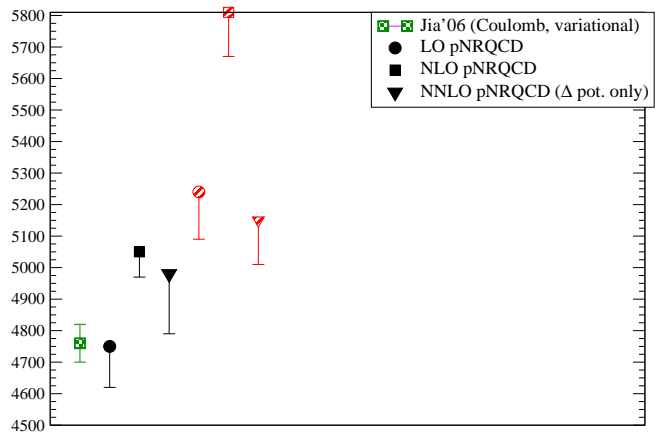}}
\caption{Several pNRQCD predictions for the $\Omega_{ccc}$ $3/2^+$ mass. The result is similar to other approaches~\cite{Flynn:2011gf}, ~\cite{Bjorken:1985ei}, etc.
}
\label{Fig:spectrum}
\end{figure}

Figure~\ref{Fig:spectrum} presents several pNRQCD predictions for the $\Omega_{ccc}$ $3/2^+$ mass. The left-most point is the estimate of Jia based on Eq.~(\ref{LOpot}). The solid, black points in the pole scheme (see below) show very reasonable agreement at leading order in spite of our much more numeric-intensive calculation. We also present the NLO and NNLO computations that are reassuringly close.  Perturbation theory seems to be working. The additional three calculations on the right are performed in the PS (potential subtraction) scheme where $V(q)\to \theta(q-\Lambda) V(q)$ is IR-truncated. Here there is a larger sensitivity. The pull-down bars correspond to the variational error (whose sign is known). 

There are no physical free parameters because we have worked out first several meson observables, so we are providing a pure prediction.

\section{Error budget and treatment of the IR}

We estimated several sources of uncertainty in our computation,  yielding a precision level of 200-250 MeV. The first is the variational approximation (a two-parameter wavefunction ansatz  overestimates binding energies in atomic physics by 25\%, or about 40 MeV in our ccc system). Also under control are the input parameters $\alpha_s$ and $m_c$, $m_b$ fit to meson observables and that contribute an additional 50 MeV to the error budget. We find that changing the NLO by the NNLO potentials amounts to some 100 MeV. 

And the largest effect, of order 200 MeV, is the treatment of the potential in the infrared. This happens because the corresponding NLO or NNLO $\alpha_s(Q^2)$ grow rapidly towards the IR in perturbation theory, and the numeric diagonalization integrates over all momenta $Q$ (analytic approaches to the meson problem do not face the infrared since the wavefunction is evaluated at scales $m\alpha_s$). One possibility is to freeze $\alpha_s$ in agreement with analytic perturbation theory or DSE's, to stay in a ``frozen''-pole mass scheme. This includes part of the IR physics but freezing is similar to a propagator resummation, and this breaks the philosophy of perturbation theory. 

Another, more drastic possibility, is to adopt the PS scheme (acknowledging that perturbation theory should not see infrared scales at all) and truncate the potential (ergo, $\alpha_s$) at some intermediate scale between, say, 600MeV-1GeV. Both possibilities have been shown in figure~\ref{Fig:spectrum}. 

Our predictions for the $QQQ$ baryon masses (in GeV) are not particularly precise, $M_{\Omega_{ccc}}=4.9(0.25)$, and $M_{\Omega_{bcc}}=8.15 (0.3)$, $M_{\Omega_{bbc}}=11.4(0.3)$, $M_{\Omega_{bbb}}=14.7(0.3)$,
but quite some improvement in the errors is possible. 

\section{Outlook and further work}

\begin{figure}[t]
\centerline{\includegraphics*[width=8.5cm]{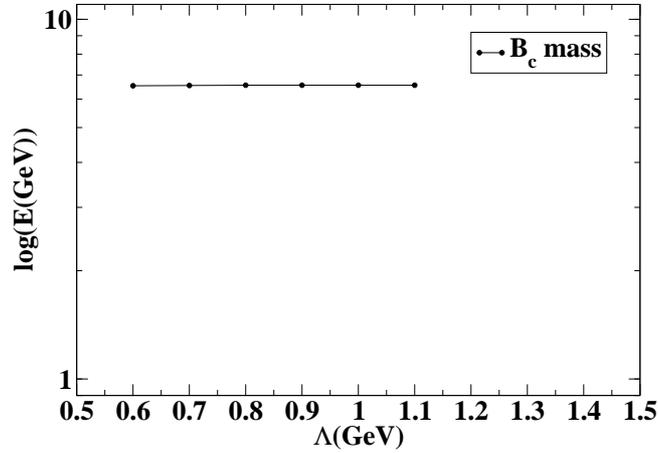}}
\caption{The $B_c$ meson mass as function of the infrared cutoff where the perturbative potential is set to zero. The simultaneous rescaling of the quark masses cancels the sensitivity to the regulating procedure and yields a stable mass prediction.}
\label{Fig:Revolution}
\end{figure}

Some of our current research focuses on assessing whether the sensitivity to the infrared cutoff~\cite{Beneke:1998rk} can be ameliorated by employing a renormalization-group equation~\cite{Hoang:2009yr}. Our result for the $B_c$ meson in figure~\ref{Fig:Revolution} is promising.
Future work should also somehow address the effect of light-quark degrees of freedom, that is not germane to pNRQCD. One can guess easily an error of order 50 MeV due to pion-cloud effects. Such ``unquenching'' of pNRQCD requires coupling its quasi-static sources to chiral perturbation theory and doing so should prove rewarding~\footnote{We thank E. Ribeiro for this observation. See~\cite{Bicudo:1998bz} for similar work combining light and heavy quark systems.}.
The generalization to many-body systems with a larger number of heavy quarks, for example $QQ\bar{Q}\bar{Q}$ tetraquarks or molecules that can populate the 7 GeV region (for charmed quarks up to the 18 GeV region (for bottom quarks) should follow similar lines, once the relevant static Wilson potentials become available.

\end{document}